\let\ifarxiv=\iftrue     
{}
{}

\newif\ifpublic\publictrue

\ifarxiv

\documentclass[12pt,a4paper]{article}

\usepackage{amsmath,amssymb}
\ifpublic\else\usepackage{showkeys}\fi
\usepackage{graphicx}
\usepackage[bookmarks=true,hyperfigures=true]{hyperref}
\usepackage[usenames, dvipsnames]{color}
\usepackage[nosort]{cite}
\usepackage[bulletsep]{collref}

\def\showkeysrefformat#1{{\normalfont\tiny\ttfamily#1}}
\makeatletter
\def\SK@@ref#1>#2\SK@{%
 {\@inlabelfalse\leavevmode\vbox to\z@{%
 \vss\SK@refcolor\rlap{\vrule\raise .75em%
  \hbox{\showkeysrefformat{#2}}}}}}
\makeatother

\usepackage[a4paper,text={160mm,247mm},centering]{geometry}

\allowdisplaybreaks[3]

\usepackage[font=small,labelfont=bf,width=0.85\textwidth]{caption}

\fi

\ifarxiv\else

\iftrue
\PassOptionsToClass{showpacs,showkeys,preprintnumbers}{revtex4-1}
\PassOptionsToClass{superscriptaddress}{revtex4-1}
\fi
\documentclass[10pt,letterpaper,twocolumn,amsmath,amssymb,final,aps,pra]{revtex4-1}

\usepackage{graphicx}
\usepackage[bookmarks=true,hyperfigures=true]{hyperref}
\usepackage[usenames, dvipsnames]{color}

\makeatletter
\let\o@a@f\@author@finish
\def\@author@finish{\o@a@f%
\let\@authors\empty\def\AF@opr##1{}\def\CO@opr##1##2##3{}%
\def\AU@opr##1##2##3{\ifx\@authors\@empty\toks@\expandafter{##2}%
  \else\toks@\expandafter{\@authors, ##2}\fi\edef\@authors{\the\toks@}}
\@AAC@list}
\makeatother

\fi

\expandafter\def\expandafter\bfseries\expandafter{\bfseries\ifmmode\else\boldmath\fi}
\expandafter\def\expandafter\mdseries\expandafter{\mdseries\ifmmode\else\unboldmath\fi}
\expandafter\def\expandafter\normalfont\expandafter{\normalfont\ifmmode\else\unboldmath\fi}

\usepackage{fixmath}

\RequirePackage[extdef]{delimset}

\let\barefrac=\frac
\renewcommand{\frac}[2]{\mathinner{\barefrac{#1}{#2}}}

\let\baresqrt=\sqrt
\makeatletter
\renewcommand{\sqrt}{\@ifnextchar[\@sqrt@space@a\@sqrt@space@b}
\def\@sqrt@space@a[#1]#2{\mathinner{\mathchoice{\mkern-3mu}{\mkern-3mu}{}{}\baresqrt[#1]{#2}}}
\def\@sqrt@space@b#1{\mathinner{\mathchoice{\mkern-3mu}{\mkern-3mu}{}{}\baresqrt{#1}}}
\makeatother

\makeatletter
\let\per@dot@old=\.
\def\.{\ifmmode\def\per@dot@sel{\mkern3mu}\else\def\per@dot@sel{\per@dot@old}\fi\per@dot@sel}
\makeatother

\newcommand{\sfrac}[2]{{\textstyle\mathord{\frac{#1}{#2}}}}
\newcommand{\half}{\sfrac{1}{2}}
\newcommand{\ihalf}{\sfrac{i}{2}}

\newcommand{\vfrac}[2]{\ifmmode\mathinner{\textstyle^{#1}\!/\!_{#2}}\else$^{#1}\!/\!_{#2}$\fi}

\newcommand{\alg}[1]{\mathfrak{#1}}
\newcommand{\Yalg}{\mathrm{Y}}
\newcommand{\grp}[1]{\mathrm{#1}}
\newcommand{\gener}[1]{\mathbb{#1}}

\newcommand{\genY}[1]{\widehat{\gener{#1}}{}}

\makeatletter\let\Re\@undefined\let\Im\@undefined\makeatother
\DeclareMathOperator{\Re}{Re}
\DeclareMathOperator{\Im}{Im}
\DeclareMathOperator{\tr}{tr}

\newcommand{\superN}{\mathcal{N}}

\newcommand{\alignbreak}[1][\nonumber]{#1\\&\mathrel{\phantom{=}}\mathord{}}

\def\[{\begin{equation}}
\def\]{\end{equation}}

\providecommand{\href}[2]{#2}
\newcommand{\arxivlink}[1]{\href{http://arxiv.org/abs/#1}{arxiv:#1}}

\makeatletter
\def\mr@ignsp#1 {\ifx\:#1\@empty\else #1\expandafter\mr@ignsp\fi}%
\newcommand{\multiref}[1]{\begingroup
\xdef\mr@no@sparg{\expandafter\mr@ignsp#1 \: }%
\def\mr@comma{}%
\@for\mr@refs:=\mr@no@sparg\do{\mr@comma\def\mr@comma{,}\ref{\mr@refs}}%
\endgroup}
\renewcommand{\eqref}[1]{(\multiref{#1})}
\makeatother

\makeatletter
\newcommand{\namedref}[2]{\hyperref[#2]{#1~\ref*{#2}}}
\newcommand{\secref}{\@ifstar{\namedref{Section}}{\namedref{Sec.}}}
\newcommand{\appref}{\@ifstar{\namedref{Appendix}}{\namedref{App.}}}
\newcommand{\tabref}{\@ifstar{\namedref{Table}}{\namedref{Tab.}}}
\newcommand{\figref}{\@ifstar{\namedref{Figure}}{\namedref{Fig.}}}
\makeatother

\let\oldbib=\thebibliography
\def\thebibliography{\phantomsection\addcontentsline{toc}{section}{\refname}\oldbib}

\let\oldtoc=\tableofcontents
\def\tableofcontents{\phantomsection\addcontentsline{toc}{section}{\contentsname}\oldtoc}

\providecommand{\hypersetup}[1]{}
\providecommand{\texorpdfstring}[2]{#1}

\hypersetup{plainpages=false}
\hypersetup{pdfpagemode=UseNone}
\hypersetup{bookmarksnumbered=true}
\hypersetup{pdfstartview=FitH}
\hypersetup{colorlinks=false}
\hypersetup{citebordercolor={.5 1 .5}}
\hypersetup{urlbordercolor={.5 1 1}}
\hypersetup{linkbordercolor={1 .7 .7}}

\makeatletter
\let\@keywords\@empty
\let\@subject\@empty
\providecommand{\keywords}[1]{\gdef\@keywords{#1}}
\providecommand{\subject}[1]{\gdef\@subject{#1}}
\def\thetitle{\@title}
\def\theauthor{\@author}
\def\thesubject{\@subject}
\def\thedate{\@date}
\def\thekeywords{\@keywords}
\makeatother
\AtBeginDocument{
\hypersetup{pdftitle={\thetitle}}%
\hypersetup{pdfauthor={\theauthor}}%
\hypersetup{pdfsubject={\thesubject}}%
\hypersetup{pdfkeywords={\thekeywords}}%
}

\ifpublic
\newcommand{\remark}[2][]{\ignorespaces}
\else
\usepackage{ifpdf}
\ifpdf\else\RequirePackage[active]{srcltx}\fi
\RequirePackage{color}

\newcommand{\remark}[2][]{{\normalfont\sffamily\hspace{1ex}\def\tmparga{#1}%
  \def\tmpargb{MR}\ifx\tmparga\tmpargb\color{magenta}\fi%
  \def\tmpargb{NB}\ifx\tmparga\tmpargb\color{blue}\fi%
  \def\tmpargb{AG}\ifx\tmparga\tmpargb\color{Orange}\fi%
  \def\tmpargb{Refs}\ifx\tmparga\tmpargb\color{Orchid}\fi%
  \def\tmpargb{}\ifx\tmparga\tmpargb\color{red}\fi%
  \def\tmpargb{}\ifx\tmparga\tmpargb\else \textbf{#1:} \fi%
  #2\hspace{1ex}}}
\fi

\newcommand{\remarkref}[1][]{{\def\tmparga{#1}\def\tmpargb{}%
  \ifx\tmparga\tmpargb\remark[Refs]{needed}\else\remark[Refs]{#1}\fi}}

\begin{document}
\title{Yangian Symmetry and Integrability\ifarxiv\\\else\space\fi 
of Planar \texorpdfstring{$\mathcal{N}=4$}{N=4} Super-Yang--Mills Theory}
\long\def\theabstract{
In this letter we establish Yangian symmetry 
of planar $\mathcal{N}=4$ super-Yang--Mills theory. 
We prove that the classical equations of motion of the model 
close onto themselves under the action of Yangian generators.
Moreover we propose an off-shell extension of our statement
which is equivalent to the invariance of the action
and prove that it is exactly satisfied. 
We assert that our relationship serves as a 
criterion for integrability in planar gauge theories
by explicitly checking that it applies to integrable ABJM theory
but not to non-integrable $\mathcal{N}=1$ super-Yang--Mills theory.
}

\ifarxiv\else
\author{Niklas Beisert}
\email{nbeisert@itp.phys.ethz.ch}
\author{Aleksander Garus}
\email{agarus@itp.phys.ethz.ch}
\affiliation{Institut f\"ur Theoretische Physik, 
Eidgen\"ossische Technische Hochschule Z\"urich, 
Wolfgang-Pauli-Strasse 27, 8093 Z\"urich, Switzerland}

\author{Matteo Rosso}
\email{matteo.rosso@physik.hu-berlin.de}
\affiliation{Institut f\"ur Physik,
Humboldt Universit\"at zu Berlin,
Zum Grossen Windkanal 6, D-12489 Berlin, Germany}

\begin{abstract}
\theabstract
\end{abstract}

\maketitle
\fi

\ifarxiv
\pdfbookmark[1]{Title Page}{title}
\thispagestyle{empty}

\begingroup\raggedleft\footnotesize\ttfamily
\arxivlink{1701.09162}\\
HU-EP-17/03
\par\endgroup

\vspace*{2cm}
\begin{center}%
\begingroup\Large\bfseries\thetitle\par\endgroup
\vspace{1cm}


\begingroup\scshape
Niklas Beisert\textsuperscript{1}, 
Aleksander Garus\textsuperscript{1} and 
Matteo Rosso\textsuperscript{2}
\endgroup
\vspace{5mm}

\textit{\textsuperscript{1}Institut f\"ur Theoretische Physik,\\
Eidgen\"ossische Technische Hochschule Z\"urich,\\
Wolfgang-Pauli-Strasse 27, 8093 Z\"urich, Switzerland}
\vspace{0.1cm}

\begingroup\ttfamily\small
\verb+{+nbeisert,agarus\verb+}+@itp.phys.ethz.ch\par
\endgroup
\vspace{5mm}

\textit{\textsuperscript{2}Institut f\"ur Physik,\\
Humboldt Universit\"at zu Berlin,\\
Zum Grossen Windkanal 6, D-12489 Berlin, Germany}
\vspace{0.1cm}

\begingroup\ttfamily\small
matteo.rosso@physik.hu-berlin.de\par
\endgroup
\vspace{5mm}
\vfill

\textbf{Abstract}\vspace{5mm}

\begin{minipage}{12.7cm}
\theabstract
\end{minipage}

\vspace*{4cm}

\end{center}

\newpage
\fi

\section{Introduction}
\label{sec:intro}

The assumption that planar $\mathcal{N}=4$ super-Yang--Mills (sYM) theory 
is integrable (see \cite{Beisert:2010jr} for a review)
has unlocked an enormous body of data 
-- not only perturbatively but also at strong and intermediate coupling --
thanks to advanced methods of integrable systems.
Two prime examples are the finite-coupling computations 
of the cusp dimension \cite{Benna:2006nd}
by means of an integral equation \cite{Beisert:2006ez}
and of the scaling dimension of the Konishi operator
\cite{Gromov:2009zb} by means of the thermodynamic Bethe ansatz
\cite{Gromov:2009tv,Bombardelli:2009ns,Arutyunov:2009ur}.
Their smooth interpolations between weak and strong coupling 
are viewed as strong confirmations of the AdS/CFT correspondence.
While the feature of integrability in this model is now 
supported by an overwhelming amount of evidence,
it largely remains a conjecture
except for certain observables in certain corners of parameter space. 
The main obstacle in proving integrability
is the lack of a proper definition for this feature
within a planar gauge theory.

A key property of integrable systems is the existence
of a large amount of hidden symmetries. 
For this model, the relevant algebra
has been identified as the Yangian $\Yalg[\alg{psu}(2,2|4)]$
\cite{Dolan:2003uh}
which is an infinite-dimensional quantum algebra based 
on the Lie superalgebra $\alg{psu}(2,2|4)$ of superconformal symmetries.
The original formulation, however, merely addressed
the spectrum of one-loop anomalous dimensions 
for which Yangian symmetry is largely broken 
due to the pertinent cyclic boundary conditions.
Sometime later, Yangian invariance has been established 
for colour-ordered scattering amplitudes at tree level \cite{Drummond:2009fd}.
Unfortunately, at loop level this symmetry is severely affected by 
infra-red singularities inherent to scattering of massless particles.
Only recently, the Yangian has been found to be a proper symmetry
of the expectation values of certain non-singular Wilson loops \cite{Muller:2013rta}.
However, the Yangian has never been shown to be  
a symmetry of the model itself.

The goal of the present letter is to fill this gap
and to establish Yangian symmetry for planar $\superN=4$ sYM.
The obvious strategy would be to show that the 
action of the model is invariant under the Yangian generators.
However, there are several difficulties in this approach:
First, the large-$N_\text{c}$ limit must play a decisive role 
for integrability is clearly restricted to the planar limit.
However, it is not a priori evident how to define 
the planar limit of the Lagrangian or of the Yangian generators.
Another difficulty lies in the apparent incompatibility 
of the Yangian with the cyclicity of the Lagrangian.
Finally, one needs to define how exactly the Yangian generators 
act on the fields.
This turns out to be a rather subtle issue within a gauge theory
because symmetries necessarily act non-linearly
in the sense that one field is mapped to a single or to several fields. 
On the one hand, non-linearity complicates the analysis, 
especially in combination with the above issues. 
On the other hand, exact non-linear invariance is 
a very strong statement which typically extends 
to the quantum theory unless the symmetry is anomalous.
The combination of these difficulties has arguably 
been a show-stopper in proving Yangian symmetry of the model in the past.
Here we start with the somewhat more moderate goal to
establish Yangian symmetry for the classical equations of motion of $\mathcal{N}=4$ sYM. 
We first show that the equations of motion close onto themselves 
under an appropriately defined action of the Yangian generators.
This alone is not sufficient to establish the Yangian as a proper symmetry
of the model. 
We therefore propose an off-shell extension of the statement 
and prove its validity.
We claim that our relationship is equivalent to Yangian invariance 
of the action, and thus serves as a suitable criterion 
for integrability in planar gauge theories.
In order to substantiate these claims, 
we consider similar superconformal gauge theory models:
on the one hand, we show that evidently non-integrable $\mathcal{N}=1$ sYM theory 
does not satisfy our relationship. 
On the other hand, we prove Yangian symmetry of 
Aharony--Bergman--Jafferis--Maldacena (ABJM) theory
\cite{Aharony:2008ug}
for which signs of integrability have been found 
following the work \cite{Minahan:2008hf}.

\section{\texorpdfstring{$\mathcal{N}=4$}{N=4} super-Yang--Mills}
\label{sec:msym}

The $\mathcal{N}=4$ sYM theory in four dimensions consists 
of the gauge field $A_\mu$, four Weyl fermions $\Psi_{a\alpha}$ 
together with their conjugates $\bar{\Psi}^a_{\dot{\alpha}}$ and six
real scalar fields $\Phi_m$.
We denote spacetime vector indices by $\mu,\nu,\ldots=0, \ldots, 3$ 
and spinor indices of the two chiralities by 
$\alpha, \beta, \ldots=1,2$ and $\dot{\alpha}, \dot{\beta},\ldots=1,2$,
respectively.
Moreover, $m,n,\ldots=1,\ldots,6$ and $a,b,\ldots=1,2,3,4$ 
denote indices for the vector and spinor
representations of $\grp{SO}(6)$, respectively.
All matter fields transform in the adjoint
representation of the gauge group. 
The covariant derivative of a generic covariant field $Z$ 
is $D_\mu Z:=\partial_\mu Z - i \comm{A_\mu}{Z}$, 
and the field strength equals 
$F_{\mu\nu}:=\partial_\mu A_\nu-\partial_\nu A_\mu- i \comm{A_\mu}{A_\nu}$.

The Lagrangian of the theory takes the form 
\begin{align}
\mathcal{L}&=
-\sfrac{1}{4}\tr F_{\mu \nu}F^{\mu \nu}
+ \tr i\bar{\Psi}^a_{\dot{\alpha}}\sigma_\mu^{\alpha \dot{\alpha}}D^{\mu}\Psi_{a \alpha}
\alignbreak
-\sfrac{i}{2} \tr \brk!{
\Psi_{a \alpha} \sigma_m^{ab} \epsilon^{\alpha \beta} \comm{\Phi^m}{\Psi_{b\beta}}
- \bar{\Psi}^a_{\dot{\alpha}} \sigma^m_{ab} 
\epsilon^{\dot{\alpha} \dot{\beta}} \comm{\Phi_m}{\bar{\Psi}^b_{\dot{\beta}}}}
\alignbreak
-\sfrac{1}{2}\tr D^\mu\Phi^m D_\mu \Phi_m 
+ \sfrac{1}{4}\tr \comm{\Phi^m}{\Phi^n}\comm{\Phi_m}{\Phi_n}.
\label{eq:lagrangian}
\end{align}
Here, $\sigma^\mu_{\alpha \dot{\alpha}}$ and $\sigma^m_{ab}$ 
denote the generalisations of the Pauli matrices to 
$(3+1)$D and to 6D, respectively.
Furthermore, $\epsilon^{\alpha \beta}$, $\epsilon^{\dot{\alpha} \dot{\beta}}$
and $\epsilon^{abcd}$ denote the totally antisymmetric tensors.
As already alluded to in the introduction, 
we are mainly going to work with the equations of motion of the theory.
To that end, we introduce a short-hand notation
for the variation of the action with respect to a generic field $Z_A$
(the indices $A,B,\ldots$ enumerate the various fields)
$\breve{Z}^A:=\delta \mathcal{S}/\delta Z_A$,
so that the equations of motion simply read $\breve{Z}^A=0$.
We have explicitly that
\begin{align}
\label{eq:N4_eom} 
\breve{\bar{\Psi}}_a^{\dot{\alpha}}
&=
i\sigma_\mu^{\alpha \dot{\alpha}}D^\mu \Psi_{a \alpha} 
+ i \epsilon^{\dot{\alpha} \dot{\gamma}} \sigma^m_{ab}
\comm{\Phi_m}{\bar{\Psi}^b_{\dot{\gamma}}},
\nonumber \\
\breve{\Psi}^{a\alpha}
& =
i\sigma_\mu^{\dot{\alpha} \alpha} D^\mu \bar{\Psi}_{\dot{\alpha}}^a 
+ i \epsilon^{\alpha \beta}\sigma_m^{ab}\comm{\Phi^m}{\Psi_{\beta b}},
\nonumber\\
\breve{\Phi}_m
&= 
D_\mu D^\mu \Phi_m 
+ \comm!{\Phi_n}{\comm{\Phi^n}{\Phi_m}} 
\ifarxiv\else\alignbreak\fi
 +\tfrac{i}{2} \sigma_m^{ab} \epsilon^{\alpha \beta} 
\acomm{\Psi_{a \alpha}}{\Psi_{b \beta}}
 + \tfrac{i}{2}  \sigma_{m,ab} \epsilon^{\dot{\alpha} \dot{\gamma}}
\acomm{\bar{\Psi}^a_{\dot{\alpha}}}{\bar{\Psi}^b_{\dot{\gamma}}},
\nonumber \\
\breve{A}_\mu &=D^\nu F_{\nu \mu}
+i \comm{\Phi_n}{D_\mu \Phi^n}
+ \sigma_\mu^{\dot{\alpha}\alpha}
\acomm{\bar{\Psi}^a_{\dot{\alpha}}}{\Psi_{a \alpha}}
\end{align}
 
$\mathcal{N}=4$ sYM theory is a superconformal theory; its action
is invariant under the four-dimensional extended superconformal algebra
$\alg{psu}(2,2|4)$, and quantum effects do not spoil such invariance.  In the
following, we will need the action of the generators of dilatations
$\gener{D}$, translations $\gener{P}_\mu$ 
and Lorentz transformation $\gener{M}_{\mu\nu}$ on a generic field $Z$ 
\begin{align}
\label{eq:poincare_action}
\gener{D} Z
&= i \brk!{x^\sigma D_\sigma + \Delta_Z } Z,%
\nonumber\\
\gener{P}^\rho Z 
&= i D^\rho Z,
\nonumber\\
\gener{L}_{\mu \nu} Z
&= i \brk!{x_\mu D_\nu - x_\nu D_\mu + \Sigma_{\mu \nu}}Z.
\end{align}
In the above equations, $\Delta_Z$ is the conformal dimension of the field $Z$, 
and $\Sigma_{\mu\nu}$ is a spin-specific
part of the transformation.
To make transformations consistent for the non-covariant gauge field $A_\mu$, 
we have to make the peculiar definitions
$\Delta_{A}:=0$ and $D_\mu A_\nu:=F_{\mu\nu}$.
Supersymmetry generators $\gener{Q}^a_\alpha$ act as 
\begin{align}
\label{eq:susy_action}
\gener{Q}^a_{\alpha} \Phi_m 
&= \sigma_m^{ab}\Psi_{b \alpha},
\nonumber \\
\gener{Q}^a_{\alpha} \Psi_{b \beta} 
&= - \half \sigma^\mu_{\beta \dot{\epsilon}}\epsilon^{\dot{\epsilon} \dot{\gamma}} 
  \sigma^{\nu}_{ \alpha \dot{\gamma}} F_{\mu \nu} \delta^a_b
+\ihalf \epsilon_{\beta \alpha} \sigma^m_{b c} \sigma_n^{c a} [\Phi_m, \Phi^n],
\nonumber \\ 
\gener{Q}^a_{\alpha} \bar{\Psi}^b_{\dot{\gamma}} 
&=i\sigma_m^{ab} \sigma^\mu_{\alpha \dot{\gamma}} D_\mu \Phi^m ,
\nonumber \\
\gener{Q}^a_{\alpha} A^\mu 
&= i\sigma^\mu_{ \alpha \dot{\gamma}}
\epsilon^{\dot{\gamma} \dot{\epsilon}}  \bar{\Psi}^a_{\dot{\epsilon}}.
\end{align}
Analogous expressions hold for 
the conjugate generator $\gener{\bar{Q}}_{a \dot{\alpha}}$, 
with the roles of $\Psi_{a \alpha}$ and $\bar{\Psi}^a_{\dot{\alpha}}$ interchanged.

\section{Yangian invariance of \texorpdfstring{$\mathcal{N}=4$}{N=4} sYM}
\label{sec:Yinv}

We want to study classical Yangian symmetry of planar $\mathcal{N}=4$ sYM;
ordinarily, one would show the invariance of the action $\mathcal{S}$. 
This works well for the superconformal generators $\gener{J}^K$
at level zero of the Yangian (the indices $K,L,\ldots$ enumerate 
a basis of the level-zero algebra $\alg{psu}(2,2|4)$),
namely $\gener{J}^K \mathcal{S}=0$.
As outlined in the introduction,
there are several difficulties in formulating invariance of the action 
to higher-level generators.
Gladly, most of them disappear when 
acting on the equations of motion instead.
In the following, we will show how this can be achieved 
and how to promote their invariance to a powerful off-shell statement.

A level-one generator $\genY{J}^K$ has a bilocal contribution
determined completely by the level-zero generators
\begin{equation}
\label{eq:jhat_bilocal}
\genY{J}^K_{\text{biloc}} = \half f^K_{MN}\. \gener{J}^M \wedge \gener{J}^N;
\end{equation}
here $f^K_{M N}$ denotes the structure constants of $\alg{psu}(2,2|4)$. 
Explicitly, the bilocal term acts in the following way 
on a sequence $Z_1Z_2\ldots Z_n$ of fields
\begin{equation}
\ifarxiv
\genY{J}^K_{\text{biloc}}\brk{Z_1\ldots Z_n}=
\fi
f^K_{MN}\sum_{1=i<j}^n Z_{1}\ldots \brk!{\gener{J}^M Z_{i}}
   \ldots \brk!{\gener{J}^N Z_{j}}\ldots Z_{n} .
\label{eq:jhat_bil_explicit}
\end{equation}
Notice that this is where the planar limit 
is relevant for Yangian symmetry:
Only in the planar limit, 
the ordering of fields within a matrix product 
is universally and unambiguously defined 
because there are no identities 
between matrix polynomials
when $N_\text{c}$ is sufficiently large.

Let us apply the simplest level-one generator $\genY{P}^\rho$,
also known as the dual superconformal generator \cite{Drummond:2009fd}, 
to the equations of motion \eqref{eq:N4_eom}. 
We will work with the easiest of them, the Dirac equation.
The bilocal part of the level-one momentum $\genY{P}^\rho$ takes the form
\begin{equation}
\label{eq:phat_bilocal}
\genY{P}^\rho_\text{biloc}
= \gener{D}\wedge \gener{P}^\rho - \gener{L}^\rho{}_\mu\wedge \gener{P}^\mu 
- \tfrac{i}{4} \sigma^{\rho,\dot{\alpha}\beta } 
\gener{\bar{Q}}_{a \dot{\alpha}}\wedge \gener{Q}^a_{\beta}.
\end{equation}
Applying it to the Dirac equation we get
\begin{align}
  \label{eq:biloc_dirac}
\genY{P}^\mu_{\text{biloc}}
\breve{\bar{\Psi}}_a^{\dot{\alpha}}
&=
-i \epsilon^{\dot{\alpha} \dot{\gamma}} \sigma^m_{ab} 
 \acomm{\Phi_m}{ D^\mu\bar{\Psi}^b_{\dot{\gamma}}}
- \ihalf \epsilon^{\dot{\alpha}\dot{\gamma}} \sigma^m_{ab} 
\acomm{D^\mu \Phi_m}{\bar{\Psi}^b_{\dot{\gamma}}} 
\alignbreak
-\ihalf \epsilon^{\dot{\alpha}\dot{\gamma}} \sigma^m_{ab} 
\sigma^\mu_{\alpha \dot{\gamma}} \sigma_\nu^{\alpha \dot{\epsilon}}
\acomm{D^\nu \Phi_m}{ \bar{\Psi}^b_{\dot{\epsilon}}} 
\ifarxiv\else\alignbreak\fi
-\ihalf \sigma^{\mu, \rho \dot{\alpha}}  
\sigma^m_{ac}\sigma_n^{cb} \acomm!{\Psi_{\rho b}}{ \comm{\Phi_m}{\Phi^n}}. 
\end{align}
It is useful to observe that the explicit $x$-dependence 
due to some of the bosonic generators in \eqref{eq:poincare_action}
drops out completely. 
This is related to the triviality of $\comm{\gener{P}^\mu}{\genY{P}^\rho}$.

A level-one generator can also have some local contributions, 
which act on a single field at a time just like the level-zero generators.
These contributions are not determined by level-zero symmetry,
and we may adjust them according to our needs.
We now ask whether there is a single-field action of $\genY{P}^\rho$
ensuring the invariance of the equations of motion.  With the following choice
for the single-field action of $\genY{P}^\mu$ 
\begin{align}
  \label{eq:phat_action}
  \genY{P}^\mu  \Phi_m &= 0 
,\nonumber\\
  \genY{P}^\mu  \Psi_{a \alpha} 
&= \half  \sigma^\mu_{\alpha \dot{\gamma}} \epsilon^{\dot{\gamma} \dot{\epsilon}} 
  \sigma^m_{a b} \acomm{ \Phi_m}{\bar{\Psi}^b_{\dot{\epsilon}}} 
,\nonumber\\
  \genY{P}^\mu  \bar{\Psi}^a_{\dot{\alpha}} 
&= \half  \sigma^\mu_{ \beta \dot{\alpha}} \epsilon^{\beta \gamma} 
  \sigma_m^{ab} \acomm{\Phi^m}{\Psi_{\gamma b}}  
,\nonumber\\
  \genY{P}^\mu  A^\rho 
&= \ihalf  \eta^{\mu \rho} \acomm{\Phi_m}{\Phi^m}
,\end{align}
the combination of local and bilocal terms in the action of
$\genY{P}$ gives 
\begin{equation}
\genY{P}^\mu \breve{\bar{\Psi}}_a^{\dot{\alpha}} 
=-\half \sigma^\mu_{\alpha \dot{\gamma}} \sigma^m_{ba}\epsilon^{\dot{\gamma} \dot{\alpha}}
 \acomm{\Phi_m}{ \breve{\Psi}^{\alpha b}}.
\label{eq:vanish_on_shell}
\end{equation}
The r.h.s.\ is proportional to $\breve{\Psi}$ and vanishes on shell.
Hence $\genY{P}^\rho$ is an on-shell symmetry 
of the $\mathcal{N}=4$ sYM Dirac equation. 

The invariance of the action $\mathcal{S}$ 
under a superconformal generator $\gener{J}^K$
implies a stronger, off-shell relationship 
for the equations of motion. 
To that end, consider the invariance of the action,
$\gener{J}^K \mathcal{S}=\breve{Z}^A\brk!{\gener{J}^K Z_A}=0$.
Now vary this equation with respect to a generic field $Z_C$
to obtain an equivalent relation which holds \emph{off-shell}
\begin{equation}
  \gener{J}^K \breve{Z}^C 
= -\breve{Z}^A\frac{\delta \brk!{\gener{J}^K Z_A}}{\delta Z_C} .
\label{eq:level0_inv}
\end{equation}
The r.h.s.\ is now a specific combination 
of the equations of motion $\breve Z^A$ given 
by the action of the generators $\gener{J}^K$ 
on the fields $Z_A$ of the theory.
Getting inspiration from the structure of the bilocal term \eqref{eq:jhat_bilocal} 
as well as from the level-zero formula \eqref{eq:level0_inv} 
we propose an analogous formula for 
the level-one Yangian generators $\genY{J}^K$:
\begin{align}
\label{eq:level1_inv}
\genY{J}^K \breve{Z}^C
&=-\breve{Z}^A\frac{\delta \brk!{\genY{J}^K Z_A}}{\delta Z_C} 
\ifarxiv\else\alignbreak\fi
+ f^K_{MN} \breve{Z}^A\brk*{\gener{J}^M \wedge \frac{\delta}{ \delta Z_C}} 
\brk!{\gener{J}^N Z_A} .
\end{align}
Let us explicitly demonstrate how the two terms 
on the r.h.s.\ are to be understood. 
The former is the direct counterpart of the r.h.s.\ of \eqref{eq:level0_inv}.
Assuming that, e.g., $\genY{J}^K Z_A=Z_1 Z_2$, 
it evaluates to 
\begin{equation}
\ifarxiv
-\breve{Z}^A\frac{\delta \brk!{\genY{J}^K Z_A}}{\delta Z_C}=
\fi
-\brk[s]!{\delta^C_1 \breve{Z}^A Z_2 + \delta^C_2 Z_1 \breve{Z}^A }. 
\label{eq:magform_local}
\end{equation}
Concerning the second term, we first 
observe that it vanishes for the linear
contribution of $\gener{J}^N$ acting on $Z_A$. 
Assuming that $\gener{J}^N Z_A=Z_1 Z_2$, it evaluates to
\begin{equation}
\ifarxiv
f^K_{MN} \breve{Z}^A\brk*{\gener{J}^M \wedge \frac{\delta}{ \delta Z_C}} 
\brk!{\gener{J}^N Z_A}=
\else
+
\fi
f^K_{MN} \brk[s]!{ \delta^C_2 \brk!{\gener{J}^M Z_1} \breve{Z}^A 
  -  \delta^C_1 \breve{Z}^A \brk!{\gener{J}^M Z_2}}.
\label{eq:magform_bilocal}
\end{equation}
It is now a straightforward exercise to verify 
that the formula \eqref{eq:level1_inv} indeed
reproduces \eqref{eq:vanish_on_shell} correctly. 
Similar checks can also be made for all the other equations of motion. 
As the supersymmetry generators $\gener{Q}^a_{\alpha}$ and $\gener{\bar{Q}}_{a\dot{\alpha}}$ 
map the Dirac equation to the remaining equations of motion, 
already the algebraic relations guarantee their invariance under $\genY{P}^\rho$. 
Most importantly, they also imply invariance 
under the remaining infinitely many Yangian generators.

We conclude that the relationship \eqref{eq:level1_inv}
holds in classical planar $\mathcal{N}=4$ sYM theory. 
Furthermore, we will demonstrate in \cite{BGRlong}, 
that it can be lifted to an invariance statement 
for the action of $\mathcal{N}=4$ sYM.
In that sense, classical planar $\mathcal{N}=4$ sYM
is invariant under Yangian symmetry.

\section{Correlation Functions}

The novel symmetry relationship \eqref{eq:level1_inv} 
should have implications for observables. 
A fundamental class of observables 
in a quantum field theory is given by correlators of fields.
Although not gauge-invariant on their own, 
they are building blocks for observables like Wilson loops, 
scattering amplitudes (via the LSZ reduction formula),
correlators of local operators or form factors. 
An ordinary symmetry results in Ward--Takahashi identities 
for correlators. 
The goal is thus to derive Ward--Takahashi identities
for level-one Yangian symmetries,
and subsequently use them to prove symmetries 
of the aforementioned gauge-invariant observables.
In the following we shall briefly sketch the level-one symmetry 
for planar correlators of the fields;
a detailed treatment can be found in the follow-up paper \cite{BGRlong}.

In order to set up the quantum theory, we should first fix the gauge 
and thus make correlation functions of the fields well-defined observables.
This step can potentially break symmetries, but gladly it leaves
the level-one Yangian symmetry intact: 
We show in \cite{BGRlong} that there exists a (trivial) extension 
of the symmetry representation on the Faddeev--Popov ghosts
such that the invariance statement \eqref{eq:level1_inv} continues to hold. 
Note that it must be supplemented by suitable terms exact under BRST symmetry
which are irrelevant for physical observables.

Ward--Takahashi identities for Yangian symmetry are obtained by acting 
with the non-linear level-one momentum $\genY{P}$ 
on some collection of fields. 
Symmetry implies that the planar quantum expectation value 
of the resulting expression vanishes. 
This amounts to some extra differential constraints 
on planar correlation functions. 
Note, however, the following few caveats:
In a gauge-fixed theory, one should expect some residual terms
due to the unphysical modes
which may violate the symmetry.
At leading order, 
these should take the form of total derivatives of the external fields.
Furthermore, the resulting identity can receive divergences at loop level. 
If the level-one symmetry can be renormalised appropriately 
to make the identity hold at higher orders in perturbation theory,
it is a proper quantum symmetry.
Otherwise, the level-one Yangian symmetry would be anomalous.

In \cite{BGRlong} we demonstrate that 
(discarding issues related to gauge fixing) 
the level-one relationship \eqref{eq:level1_inv} 
combined with superconformal symmetry results 
in additional Ward--Takahashi identities at tree level 
at least up to four external points. 
Therefore Yangian symmetry indeed has concrete implications 
for planar correlators.

\section{Other superconformal theories}

It is tempting to consider
the relationship eq.~\eqref{eq:level1_inv}
as a criterion for integrability in 
other planar gauge theories.
The criterion can be applied to all models 
whose global symmetries form a semi-simple Lie (super)algebra.
In the following we support our proposal by 
showing that the criterion agrees with our expectations 
for two sample gauge theories with classical superconformal symmetry.

First we consider $\superN=1$ pure sYM theory in four dimensions
which is non-integrable.
This theory is classically superconformal, its global
symmetry algebra being $\alg{su}(2,2|1)$; it contains a single vector field and
a single Weyl fermion. The Lagrangian of this theory can be
obtained from $\superN=4$ sYM by dropping all the scalar fields and
retaining one single Weyl fermion; it reads
\begin{equation}%
  \label{eq:n1_action}
  \mathcal{L}=-\sfrac{1}{4}\tr F_{\mu \nu}F^{\mu \nu} 
    + \tr i\bar{\Psi}_{\dot{\alpha}}\sigma_\mu^{\dot{\alpha} \beta}D^\mu \Psi_{\beta}.
\end{equation}
Similarly, the action of the superconformal generators can be 
obtained from the corresponding expressions for $\superN=4$ sYM,
eq.~\eqref{eq:poincare_action,eq:susy_action}.  

As before, we can check our criterion for Yangian symmetry
using the level-one generator $\genY{P}^\rho$ of
$\Yalg[\alg{su}(2,2|1)]$ acting on the Dirac equation of this theory. 
Considering the quantum numbers, 
one finds no admissible terms for the single-field action of $\genY{P}^\rho$,
\begin{equation}
  \label{eq:n=1_phat}
  \genY{P}^\rho A_\mu = \genY{P}^\rho \Psi_{\alpha} 
= \genY{P}^\rho \bar{\Psi}_{\dot{\beta}}=0.
\end{equation}
The computation of the l.h.s.\ of eq.~\eqref{eq:level1_inv} in this case yields a
term proportional to $\acomm{\bar{\Psi}_{\dot{\alpha}}}{F_{\mu\nu}}$, whereas
the r.h.s.\ vanishes identically; 
in $\superN=4$ sYM the latter is non-trivial and cancels the former 
as well as all other arising terms. 
Our criterion therefore states that $\superN=1$ sYM does not possess Yangian symmetry, 
in accordance with the fact that this theory, as a whole, is not integrable.
Importantly, this demonstrates that it is not merely 
an artefact due to other properties such as gauge or superconformal symmetry.

The second example we study is the so-called ABJM theory, 
a Chern--Simons-matter theory with gauge group $\grp{U}(N_\text{c})\times \grp{U}(N_\text{c})$.
The matter sector consists of four chiral multiplets $\Phi^a$, $\Psi_{a \alpha}$
transforming in the bifundamental of the
gauge groups, together with their conjugate multiplets $\bar{\Phi}_a$,
$\bar{\Psi}{}^a_\alpha$ transforming in the conjugate bifundamental; 
the gauge fields associated with the two components of the gauge group 
are $A_\mu$ and $\tilde{A}_\mu$. 
ABJM is a superconformal theory with $\mathcal{N}=6$
supersymmetry, its global symmetry algebra being $\alg{osp}(6|4)$. 
Similarly to $\superN=4$ sYM, 
this theory appears to be exactly integrable in the planar limit, 
and also in this case its integrability is encoded in the existence of a
Yangian symmetry \cite{Bargheer:2010hn} based on its superconformal algebra.

In order to study the Yangian symmetry of planar ABJM, we start from the Lagrangian
\begin{equation}
    \mathcal{L}= %
     \mathcal{L}_{\text{CS}}-\mathcal{L}_{\widetilde{\text{CS}}}
    + \bar{\Psi}{}^{a \alpha} D_\mu \gamma^\mu_{\alpha\beta} \Psi_{a}^\beta
    + D_\mu \Phi^a D^\mu \bar{\Phi}_a+\ldots.
      \label{eq:abjmlag}
\end{equation}
Here $\mathcal{L}_{\text{CS}}$ and $\mathcal{L}_{\widetilde{\text{CS}}}$ 
are the Chern--Simons terms, 
and we omitted the Yukawa couplings and scalar potential.

We want to check whether eq.~\eqref{eq:level1_inv} holds for the Dirac equation
of ABJM
\begin{align}
\breve{\bar{\Psi}}_{a \alpha}
&=
- \gamma^\mu_{\alpha\gamma} \epsilon^{\gamma\beta} D_\mu \Psi_{a \beta} %
+\Psi_{a \alpha} \bar{\Phi}{}_{b} \Phi^b 
- \Phi_b \bar{\Phi}{}^b \Psi_{a \alpha}
\alignbreak
+2 \Phi^b \bar{\Phi}{}_a \Psi_{b \alpha} 
 - 2\Psi_{b \alpha} \bar{\Phi}{}_a \Phi^b 
 +2 \epsilon_{a b c d} \Phi^b \bar{\Psi}{}^{c}_\alpha \Phi^d
.\label{eq:dirac_abjm}
\end{align}
As before, we consider the level-one generator $\genY{P}^\rho$ of
$\Yalg[\alg{osp}(6|4)]$, whose bilocal part is of the same form described
in eq.~\eqref{eq:phat_bilocal},
\begin{equation}
  \label{eq:phat_abjm}
    \genY{P}^\rho_\text{biloc}
= \gener{D}\wedge \gener{P}^\rho - \gener{L}^\rho{}_\mu\wedge \gener{P}^\mu %
  + \tfrac{1}{16}\epsilon^{a b c d} \gamma^{\rho,\alpha\beta } 
  \gener{Q}_{a b \,\alpha}\wedge \gener{Q}_{c d \, \beta}.
\end{equation}
In order to compute the action of $\genY{P}^\rho$ on
$\breve{\bar{\Psi}}_{a \alpha}$, 
we need the action of the supercharges on the fields 
\begin{align}
\label{eq:abjm_q}
\gener{Q}_{a b \, \alpha}  \Phi^c 
&=
\delta^c_a \Psi_{b\alpha} - \delta^c_b \Psi_{a\alpha},
\nonumber\\
\gener{Q}_{a b \, \alpha}  \Psi_{c\beta} 
&= 
  -\epsilon_{a b c d} \gamma^\mu_{\alpha\beta} D_\mu \Phi^d
    - 2 \.  \epsilon_{\alpha\beta} \epsilon_{a b d e} \Phi^d \bar{\Phi}_c \Phi^e
\ifarxiv\else\alignbreak\fi
-  \epsilon_{\alpha\beta} \epsilon_{a b c d}\brk!{\Phi^d \bar{\Phi}_e \Phi^e 
- \Phi^e \bar{\Phi}_e \Phi^d},
\\ \nonumber
\gener{Q}_{a b \,\alpha}  A^\mu
&=
2 \gamma^\mu_{\alpha\beta}\. \epsilon^{\beta\gamma} \brk!{\Psi_{b \gamma}\bar{\Phi}_a 
- \Psi_{a \gamma}\bar{\Phi}_b +  \epsilon_{a b c d} \Phi^c \bar{\Psi}^d_{\gamma}},%
\end{align}
and similarly for $\bar{\Phi}_a$, $\bar{\Psi}^{a}_\alpha$ and $\tilde{A}_\mu$.
We can now show that eq.~\eqref{eq:level1_inv} holds for $\genY{P}^\rho$ acting
on $ \breve{\bar{\Psi}}_{a \alpha}$, provided that the single-field action of
$\genY{P}^\rho$ is
\begin{align}
  \genY{P}^\rho  \Phi^a &=0, 
\nonumber\\
  \genY{P}^\rho  \Psi_{a \alpha}&= %
  \gamma^\rho_{\alpha\beta}\epsilon^{\beta\gamma} \bigl[
  \Psi_{a\gamma} \bar{\Phi}{}_b \Phi^b + \Phi^b \bar{\Phi}_b \Psi_{a\gamma}
\ifarxiv\else\alignbreak\fi
  -2 \Psi_{b\gamma} \bar{\Phi}_a \Phi^b - 2 \Phi^b \bar{\Phi}_a \Psi_{b\gamma}%
  \bigr],
\nonumber \\
  \genY{P}^\rho  A^\mu&=%
  \tfrac{1}{2} \epsilon^{\rho\mu\nu} \brk!{D_\nu\Phi^a\bar{\Phi}_a %
                        + \Phi^a D_\nu\bar{\Phi}_a} 
\ifarxiv\else\alignbreak\fi
  + 2 \eta^{\mu\rho} %
  \brk!{\Phi^a \bar{\Phi}_b \Phi^b \bar{\Phi}_a 
  - \epsilon^{\alpha\beta}\Psi_{a \alpha}
  \bar{\Psi}{}^{a}_\beta},
\label{eq:abjm_phatsf}
\end{align}%
and similarly for $\bar{\Phi}_a$, $\bar{\Psi}^{a}_\gamma$ and
$\tilde{A}_\mu$. It is similarly possible to show that the off-shell invariance
condition eq.~\eqref{eq:level1_inv} holds for all the equations of motion of
ABJM, and that therefore planar ABJM is classically Yangian invariant.
However, the compatibility with gauge fixing, 
cf.\ the comments in the previous section, 
is subtler than for $\superN=4$ sYM.
At the moment we do not understand how to compensate terms 
due to the non-linear action eq.~\eqref{eq:abjm_q} 
of the supercharges on the gauge fields.
We will return to this issue in \cite{BGRlong}.

\section{Comments and conclusions}
\label{sec:conclusions}

In this letter we showed that the equations of motion of both
$\mathcal{N}=4$ super-Yang--Mills and ABJM -- 
two superconformal field theories which are apparently integrable in the planar limit
-- are invariant under the Yangian of the relevant superconformal algebra. 
Moreover, we derived an \emph{off-shell} relationship \eqref{eq:level1_inv}
which serves as a formal statement of invariance
of the classical theory.
It can therefore be seen as a criterion for integrability in planar gauge theories.

In a companion paper \cite{BGRlong} we will provide a more detailed
account of our claims and elaborate on some more advanced aspects
which we can merely touch upon in the present letter:
Starting from \eqref{eq:level1_inv}
we will derive a definition for Yangian invariance of the action 
of a planar gauge theory.
Furthermore, we will show how Yangian symmetry survives gauge fixing.
Finally, we will establish novel Ward--Takahashi identities
for planar correlators of fields.

An important next goal is to analyse the validity of Yangian symmetry 
at the quantum level. In other words, do the Ward--Takahashi identities 
continue to hold, at least at the level of loop integrands?
And does renormalisation and performing the loop integrations
render the symmetry anomalous?
Our strong expectation is that the symmetry will hold exactly 
at all values of the coupling constant
because we know that integrability leads to 
consistent results at weak, intermediate and strong coupling, 
see \cite{Beisert:2010jr,Benna:2006nd,Beisert:2006ez,Gromov:2009zb,%
  Gromov:2009tv,Bombardelli:2009ns,Arutyunov:2009ur}.
Note that, likewise, the AdS/CFT dual string theory model 
is integrable \cite{Bena:2003wd}. 
However, it is to be expected that particular observables, 
e.g.\ Wilson loops with cusps, 
will break Yangian symmetry to some extent.
Therefore, it will be most important to apply
our definition of Yangian symmetry to observables
and to understand where and how it is broken concretely.


\pdfbookmark[1]{Acknowledgements}{ack}
\section*{Acknowledgements}

The authors would like to thank 
N.\ Drukker
and 
J.\ Plefka
for interesting discussions. 

The work of NB and AG is partially supported by grant no.\ 615203 
from the European Research Council under the FP7
and by the Swiss National Science Foundation
through the NCCR SwissMAP.
The work of MR is supported by the grant PL 457/3-1 
``Yangian Symmetry in Quantum Gauge Field Theory'' 
of the German Research Foundation.
 

\ifarxiv
\bibliographystyle{nb}
\fi
\bibliography{N4Yang} 

\end{document}